\documentclass[twocolumn,aps,superscriptaddress,prL,floatfix,showpacs,longbibliography]{revtex4-2}

\usepackage{graphicx}
\usepackage{dcolumn}
\usepackage{bm}

\usepackage{ifthen,amsmath,amssymb}
\usepackage{hyperref}
\usepackage{float}
\usepackage{aas_macros}
\usepackage[export]{adjustbox}
\usepackage{bookmark}
\usepackage{newtxtext,newtxmath}
\usepackage[dvipsnames,table]{xcolor}

\newcommand{\bx}{\boldsymbol x}
\newcommand{\bvel}{\boldsymbol v}
\newcommand{\bn}{\boldsymbol n}

\begin{document}

\preprint{APS/123-QED}

\title{Mapping the gas density with the kinematic Sunyaev-Zel'dovich and patchy screening effects: a self-consistent comparison}

\author{Boryana Hadzhiyska}
\email{boryanah@alumni.princeton.edu}
\affiliation{Miller Institute for Basic Research in Science, University of California, Berkeley, CA, 94720, USA}
\affiliation{Physics Division, Lawrence Berkeley National Laboratory, Berkeley, CA 94720, USA}
\affiliation{Berkeley Center for Cosmological Physics, Department of Physics, University of California, Berkeley, CA 94720, USA}

\author{Noah Sailer}
\affiliation{Berkeley Center for Cosmological Physics, Department of Physics, University of California, Berkeley, CA 94720, USA}
\affiliation{Physics Division, Lawrence Berkeley National Laboratory, Berkeley, CA 94720, USA}

\author{Simone Ferraro}
\affiliation{Physics Division, Lawrence Berkeley National Laboratory, Berkeley, CA 94720, USA}
\affiliation{Berkeley Center for Cosmological Physics, Department of Physics, University of California, Berkeley, CA 94720, USA}


\date{\today}

\begin{abstract}

The secondary anisotropies of the cosmic microwave background (CMB) provide a wealth of astrophysical and cosmological information. Pairing measurements of the CMB temperature map obtained by DR5 of the Atacama Cosmology Telescope (ACT) with the large-scale structure imaging survey conducted by the Dark Energy Spectroscopic Instrument, DECaLS DR9, we investigate two effects that are sensitive
to the gas density $\tau$: kinematic Sunyaev-Zel'dovich (kSZ) and `patchy screening' (also known as `anisotropic screening'). 
In particular, we measure the stacked profiles around Luminous Red Galaxies (LRGs) at a mean redshift of $z \approx 0.7$. We detect the kSZ signal at 7.2$\sigma$, and we find a signal at $\sim 4.1\sigma$ for the patchy screening estimator, which is in excess relative to the kSZ signal. We attribute this excess to contamination from CMB lensing. We demonstrate the effect of lensing using $N$-body simulations, and we show that the screening signal is dominated by it. Accounting for lensing, our measurement places a 95\% upper bound on the optical depth of the Extended DESI LRG sample of $\tau <$ 2.5 $10^{-4}$ for a mean value of the sample of $\tau \approx$ 1.6 $10^{-4}$.
Furthermore, via hydro simulations, we show that the underlying optical depth signal measured by both effects (after removing the CMB lensing contribution) is in perfect agreement when adopting either a Compensated Aperture Photometry (CAP) filter or a high-pass filter.
Consistent with previous measurements, we see evidence for excess baryonic feedback around DESI LRGs in the patchy screening measurement. 
Provided both effects are measured with high signal-to-noise, one can measure the amplitude ratio between them, which is proportional to the root-mean-square velocity of the host halo sample, and place constraints on velocity-sensitive models such as modified gravity and phantom dark energy.

\end{abstract}

\maketitle


\section{Introduction}
\label{sec:intro}

A long-standing pursuit in astrophysics, mapping the distribution of baryons (gas) in the Universe holds the key to unraveling the formation and thermodynamic evolution of galaxy groups and clusters \citep{Ostriker_2005,Nagai_2007,McNamara_2007,Battaglia_2010,Moser_2022}. Moreover, their elusive connection to the underlying dark matter poses a significant challenge for cosmological analyses such as weak lensing studies, for which the effect of baryons remains one of the most major unknowns \citep{Amon_2022,Arico_2023}. Thus, inventorying the gas in low-density regions, known as the warm-hot intergalactic medium (WHIM), not only aids in resolving the so-called ``missing baryon'' problem \citep{1992Persic, Fukugita_2004, 2012Shull, Bullock2017, Battaglia_2017}, but also lets us disentangle astrophysical effects from manifestations of exotic cosmological models.

While X-ray studies have successfully characterized the properties of gas near the centers of massive clusters 
\citep[e.g.,][]{McNamara_2007,Ruan_2015,Crichton_2016,deGraaff_2019,Eckert_2019,Zenteno_2020,Pandey_2022}, doing so for lower-mass systems and for gas located on the outskirts of groups and clusters has proved a more difficult task. Among the most powerful tools for studying the properties of gas is the measurement of the Sunyaev-Zel'dovich (SZ) effect, which comes in two flavors. The thermal SZ (tSZ) effect stems from the inverse Compton scattering of cosmic microwave background (CMB) photons by the hot thermal gas within groups and clusters. Its magnitude, directly proportional to the electron pressure integrated along the line-of-sight, offers insights into the thermodynamic conditions of these cosmic structures, especially close to their hot and dense centers. In contrast, the kinematic SZ (kSZ) effect results from the encounters between CMB photons and free electrons moving in bulk relative to the CMB rest-frame. Thus, the kSZ effect is well-suited for tracing the spatial distribution of baryons, even in the outskirts of galaxy groups and clusters (e.g., see \citep{1999PhR...310...97B,2019SSRv..215...17M} for a review on the SZ effect).

Another potent probe for detecting the gas in groups and clusters is the ``patchy screening'' (also sometimes referred to as ``anisotropic screening'') effect, resulting from the Thomson scattering of CMB photons off free electrons on their path from the surface of last scattering to the observer. This leads to a characteristic damping of the primary CMB anisotropies that is spatially anisotropic \citep{Ostriker_1986,Vishniac_1987,Hu_1994,Dodelson_1995,Persi_1995,Gruzinov_1998,Dvorkin_2009a,Dvorkin_2009}. On small scales of order arcminutes, these anisotropies are proportional to the product between the optical depth integrated along the line-of-sight and the primary CMB temperature fluctuations.

\begin{equation}
    \left( \frac{\Delta T}{T_{\rm CMB}} \right)_{\rm kSZ} \approx - \tau_{\rm gal} \left( \frac{v_r}{c} \right)
\end{equation}
where $\tau_{\rm gal} = \int a n_e \sigma_T d\chi$ is the optical depth of the galaxy in question and $v_r$ is the radial velocity.
\begin{equation}
    \left( \frac{\Delta T}{T_{\rm CMB}} \right)_{\rm kSZ, rms} \sim 0.1 \mu{\rm K} \left( \frac{M_{\rm vir}}{10^{13} M_\odot/h}\right)\left( \frac{v_r^{\rm rms}}{300 {\rm km/s}}\right)
\end{equation}

\begin{equation}
    \left( \frac{\Delta T}{T_{\rm CMB}} \right)_{\rm scr} \approx - \tau_{\rm gal} \left( \frac{\Delta T}{T_{\rm CMB}} \right)_{\rm primary}
\end{equation}
Since for the primary CMB the rms fluctuation is $(\Delta T)_{\rm rms} \approx 110 \mu$K, we obtain
\begin{equation}
    \left( \frac{\Delta T}{T_{\rm CMB}} \right)_{\rm scr, rms} \sim 0.04 \mu{\rm K} \left( \frac{M_{\rm vir}}{10^{13} M_\odot/h}\right)
\end{equation}

Since velocity evolves slowly with redshift and is always close to $\sim 300$ km/s at the redshifts of interest, we conclude that the patchy signal is always a factor of $\approx 24$ smaller than kSZ for any galaxy mass.

Both the kSZ and the patchy screening effects are prominent during the epoch of reionization, when the cosmic ionization fraction experiences large spatial fluctuations, and in the late Universe, when structure collapses under the influences of gravity and leads to large peaks in the electron density \citep[e.g.][]{Roy_2018,Feng_2019,Namikawa_2021,Roy_2022}. The benefit of both probes is that they are directly proportional to the electron (gas) density along the line-of-sight and unlike X-rays and the tSZ have no dependence on other thermodynamic quantities, yielding a direct measure of the projected baryon profiles of galaxy groups and clusters. Because their weightings are different: the kSZ effect depends on the line-of-sight velocity, whereas patchy screening depends on the size of the primary CMB anisotropy, these probes are highly complementary and give us a way of validating conclusions about the gas distribution, calibrating the effect of systematics such as reconstruction noise and CIB/tSZ contamination, and estimating the characteristic line-of-sight velocities of galaxy groups and clusters. Deriving the velocity field is instrumental for many cosmological applications, as it can allow us to stress test modified gravity models and differentiate between different dark energy scenarios. Additionally, on very large scales, $ k \lesssim 0.01$, primordial non-Gaussianity models of the local-type predict a scale-dependent signal, which can be measured with much higher accuracy if one has access to the joint tracer density and velocity fields \cite{2019PhRvD.100h3508M}.

Measuring the late-time kSZ and the patchy screening signals requires access to a high-resolution, large-area CMB experiment and a large-scale structure (LSS) survey. Ground based CMB experiments, such as the Atacama Cosmology Telescope (ACT), South Pole Telescope, the Simons Observatory, CMB-HD, and CMB-S4 \citep{Benson_2014,Henderson_2016,SO_2019,2022arXiv220305728T,S4_2016}, are thus suitable on small scales and can be complemented by low-resolution, large-area CMB surveys such as \textit{Planck} and \textit{WMAP} \citep{planck2016-l01,bennett2012}. To bring these two effects into light, we typically cross-correlate the CMB temperature map with a galaxy survey, such as the Baryon Oscillation Spectroscopic Survey (BOSS) \cite{2013_Dawson, Alam_2017}, the Dark Energy Survey (DES) \cite{Abbott_2022}, and the Dark Energy Spectroscopic Instrument \cite{2016arXiv161100036D}.

Over the years, a number of different measurements of the kSZ effect have been performed using a variety of techniques such as the pairwise method \cite{Hand:2012ui, Bernardis_2017, Soergel_2016, Sugiyama_2018, Calafut_2021, li2024detection}, projected fields \cite{Hill_2016, Ferraro_2016, Kusiak_2021, Bolliet_2022}, matched filter optimization \cite{Li_2014}, and the velocity reconstruction stacking method \cite{Schaan:2015uaa,Schaan2021,Tanimura_2021,Mallaby-Kay_2023,2024MNRAS.534..655B,2024arXiv240707152H}. Recently, Refs.~\cite{2024arXiv240113033C,2024PhRvD.109j3539S} performed a stacked measurement of the patchy screening effect using ACT and unWISE data \citep{Schlafly_2019,Krolewski_2020,Krolewski_2021}
and put an upper bound on the optical depth. 

In this work, we make the first self-consistent measurement of the kSZ and patchy screening effects on the same galaxy sample, provided by the DESI imaging survey, in an effort to test and validate both methods independently as well as place constraints on the velocity field along the line-of-sight to which the kSZ effect is sensitive. This paper is organized as follows. In Section~\ref{sec:methods}, we describe the CMB experiment ACT and the LSS survey DESI before proceeding to discuss the techniques used for measuring the kSZ and patchy screening effects on ACT and DESI data. In Section~\ref{sec:results}, we summarize our main results of comparing the baryon profiles derived using both probes and comment on their consistency and the benefits of having access to the velocity field. Finally, in Section~\ref{sec:conclusions} we present some concluding remarks and discuss the future of joint CMB and LSS analysis.


\section{Data}

In this section, we summarize the observational and simulation data sets used in this study.

\subsection{Dark Energy Spectroscopic Instrument}

The Dark Energy Spectroscopic Instrument is a robotic, fiber-fed, highly multiplexed spectroscopic telescope that operates on the Mayall 4-meter telescope at Kitt Peak National Observatory \citep{2022AJ....164..207D}. DESI can obtain simultaneous spectra of almost 5000 objects over a $\sim$$3^\circ$ field \citep{2016arXiv161100037D,2023AJ....165....9S,2023arXiv230606310M} and is currently conducting a five-year dark energy survey of about a third of the sky \citep{2013arXiv1308.0847L}. The end product will comprise spectra for approximately 40 million galaxies and quasars \citep{2016arXiv161100036D}.


Here, we use the Extended photometric sample of Luminous Red Galaxies (LRGs) 
\cite{2023AJ....165...58Z,Zhou:2023gji} from the DESI Legacy Imaging Survey, which was used to select DESI targets from three telescopes: Blanco for Dark Energy Camera Legacy Survey (DECaLS), Mayall for the Mayall $z$-band Legacy Survey (MzLS), and Bok for the Beijing–Arizona Sky Survey (BASS). We also employ the photometric redshifts presented in \citet{Zhou:2023gji} for Data Release 9 (DR9). 



\subsection{Atacama Cosmology Telescope}

This paper employs the f090 and f150 single-drequency maps \cite{ACT:2023wcq,2020JCAP...12..046N} from Data Release 5 (DR5) of the Atacama Cosmology Telescope (ACT), a 6m telescope that was located in the Atacama Desert of Chile and measured the CMB from 2007 to 2022. The DR5 data comprise multifrequency observations from 2017 to 2022 covering roughly a third of the sky at three frequency bands: f090 (77–112 GHz), f150 (124–172 GHz), and f220 (182–277 GHz), and uses \textit{Planck} data on large scales \citep{2016ApJS..227...21T,2024ApJ...962..113M}. The observational program of DR5 targeted the `wide' field. 
For this work, we use only the night-time portion of the data taken in the first five observing seasons (2017-2021).
The ACT maps are produced in the plate-carr\'{e}e (CAR) projection scheme. This analysis uses the first version of the ACT DR5 maps, dr5.01. 

We apply a mask to the ACT maps that effectively discards all galaxies within 10 arcmin of a point source or a cluster \citep{2024ApJ...962..112Q}. In addition, we remove galaxies around which the measured filtered temperature decrements are more than a $5\sigma$ outlier in any of the bins. That way, we minimize the impact of especially bright objects and avoid introducing bias to the stacked profiles.


\section{Methods}
\label{sec:methods}

In this section, we introduce two methods for measuring the baryon profiles around DESI galaxies using ACT data: stacking of the kSZ effect and stacking of the patchy screening effect. We make use of the publicly available pipeline \texttt{ThumbStack} \footnote{\url{https://github.com/EmmanuelSchaan/ThumbStack}}.

\subsection{Filtering of the CMB map}
\label{sec:filter}

In this work, we make use of a high-pass filtered map for the kSZ stacking measurement and both a high-pass as well as a low-pass filtered map for the patchy screening measurement, following v1 of Ref.~\cite{2024arXiv240113033C}.

To obtain the small-scale temperature map, we apply the following filter in harmonic space, $f^\mathrm{hi}_\ell$,
\begin{align}
    f^\mathrm{hi}_\ell=\begin{cases}
			0 & \text{if $\ell<2350$}\\
            \sin\left(\frac{(\ell-2350)\pi}{300}\right) & \text{otherwise} \\
            1 &  \text{if $\ell>2500$}.
		 \end{cases}
\label{eq:high_pass_filter}
\end{align}

For the large-scale temperature map used in the patchy screening measurement, we adopt the filter $f^\mathrm{lo}_\ell$,
\begin{align}
    f^\mathrm{lo}_\ell=\begin{cases}
			1, & \text{if $\ell<2000$}\\
            \cos\left(\frac{(\ell-2000)\pi}{300}\right) & \text{otherwise} \\
            0 &  \text{if $\ell>2150$}.
		 \end{cases}
\end{align}

The cutoff scales are chosen to be non-overlapping so that the two maps are not strongly correlated with each other, and the smooth roll-off is applied so as to reduce the ringing from the harmonic space transforms. The location of the cut, $\ell \sim 2000$, ensures that the large-scale temperature sign is preserved and simplifies the form of the estimator by eliminating the need to include an inverse-variance filter.


\subsection{Stacking of the kSZ effect} 

To obtain the stacked profiles of LRGs through the kSZ effect, we need two ingredients: an estimate of the line-of-sight velocity, and a measurement of the CMB temperature decrement at the location of each galaxy.

We measure the CMB temperature decrements, $\mathcal{T}_i(\theta_d)$, around each galaxy $i$ using a `meanring' aperture photometry filter defined as:
\begin{equation}
\label{eq:ap}
\mathcal{T}(\theta_d) = \mathcal{N}^{-1}(\theta_d)
\int_{\theta_j}^{\theta_{j+1}} {\rm d}^2\theta \, T^{\rm hi}(\theta) ,
\end{equation}
where $T^{\rm hi}(\theta)$ is the high-pass filtered CMB temperature map, and the normalization is chosen as:
\begin{equation}
\mathcal{N}(\theta_d) = \int_{\theta_j}^{\theta_{j+1}} {\rm d}^2\theta .
\end{equation}
In other words, we measure the mean temperature fluctuation in a ring with inner radius $\theta_j$ and an outer radius $\theta_{j+1}$ at each radial bin, $d$. Similarly to Ref.~\cite{2024arXiv240113033C}, we choose 9 radial bins (i.e., $j = 0 \cdots 9$), spanning between 0 and 10 arcmin. For the LRGs, 1 arcmin corresponds roughly to $0.5 \ {\rm Mpc}/h$ ($\bar z \approx 0.7$). Since we obtain the measurement in several radial bins, we have sufficient information to reconstruct the projected gas profiles unlike the case of using a single bin for the aperture photometry measurement or a matched filter.

The other commonly used filter, which has been adopted for several kSZ analyses (see e.g., Refs.~\cite{Schaan_2021,2024arXiv240707152H}), is the compensated aperture photometry (CAP) filter. Similarly to the `meanring' filter, we vary the aperture radius $\theta_d$, measuring the temperature fluctuation $\mathcal{T}(\theta_d)$ as:
\begin{equation}
  \mathcal{T}(\theta_d) =
\int_0^{\theta_d} d^2\theta \hspace{0.05 cm} \delta T(\theta)
- \int_{\theta_d}^{\sqrt{2} \theta_d} d^2\theta \hspace{0.05 cm} \delta T(\theta),
    \label{eq:CAP}
\end{equation}
where $\delta T$ are the temperature decrements measured around each galaxy on the unfiltered CMB temperature map.

The second necessary ingredient is an estimate (reconstruction) of the velocity field, which we infer by solving the linearized continuity equation in redshift space:
\begin{equation}
    \nabla \cdot \bvel + \frac{f}{b} \nabla \cdot [(\bvel \cdot \hat \bn) \hat \bn] = -a H f \frac{\delta_g}{b}
\end{equation}
where $\delta_g$ is the observed galaxy number overdensity, $H(z)$ is the redshift-dependent Hubble parameter, $f$ is the logarithmic growth rate, defined as $f \equiv d \ln(D)/d \ln(a)$ with $D(a)$ the growth factor and $a$ the scale factor. Here, we assume that the galaxy overdensity $\delta_g$ is related to the matter overdensity in redshift space, $\delta$, by a linear bias factor, $b$, such that $\delta_g = b \delta$.

To estimate the individual galaxy velocities, we adopt the standard reconstruction method adopted by many BAO analyses, which yields the first-order galaxy displacement field and can be converted into an estimate for the velocity via:
\begin{equation}
  \bvel^{\rm rec}(\bx) = f(z) a(z) H(z) \psi(\bx) .
\end{equation}
In principle, this gives us access to the 3D reconstructed velocity field, but the component that matters for kSZ analysis is only the line-of-sight one, which we denote by $v_\parallel^{\rm rec} \equiv v^{\rm rec}$. In particular, we apply the \texttt{MultiGrid} implementation to solve the linearized system of partial differential equations via the package `pyrecon'\footnote{\url{https://github.com/cosmodesi/pyrecon}} \cite{2015MNRAS.450.3822W}. 

Once we have attained the reconstructed line-of-sight velocity and temperature decrements for each galaxy, we apply the velocity-weighted, uniform-mean estimator from \cite{Schaan21}:
\begin{equation}
    \hat{\tau}_{\rm kSZ}(\theta_d) = -
    \frac{1}{r}
    \frac{v_{\rm rms}^{\rm rec}}{c}
    \frac{\sum_i {T}_i^{\rm hi}(\theta_d) (v_{{\rm rec}, i}/c)}{\sum_i (v_{{\rm rec}, i}/c)^2} T_{\rm CMB},
    \label{eq:kSZ_est}
\end{equation}
where the sum is over all galaxies, $T_{\rm CMB} = 2.7255$ is the mean CMB temperature, $v_{\rm rms}^{\rm rec}$ is the rms of the radial component of the reconstructed velocities, $v_{{\rm rec}, i}$, $c$ is the speed of light, and $r \equiv \langle v^{\rm rec} v^{\rm true} \rangle/ (v_{\rm rms}^{\rm rec} v_{\rm rms}^{\rm true})$ is the cross-correlation coefficient between the reconstructed and true velocity, with the averaging being done over all galaxies. Typically \cite{Schaan_2021,2024arXiv240707152H}, this coefficient is evaluated from mock simulations, and the value Ref.~\cite{2024PhRvD.109j3534H} find for a photometric survey with DESI-like specifications and some cleaning of the redshift outliers is $r = 0.3 \pm 0.03$. If this normalization is not applied, the measured profiles and their covariance will be underestimated by $1/r$ and $1/r^2$, respectively, relative to the true optical depth. 

Similarly to Ref.~\cite{2024arXiv240707152H}, we 
ensure that the number of galaxies in each velocity bin is symmetric around the mean by random downsampling, which avoids unwanted biases from massive clusters and guarantees that in the absence of a kSZ signal, our estimator yields zero mean signal. Since we sum over an equal number of galaxies with a positive and a negative sign per absolute value bin in the velocity, additive contaminants such as the cosmic infrared background (CIB) and the tSZ effect cancel, as has been demonstrated in Refs.~\cite{Schaan_2016,Schaan_2021,2024arXiv240707152H}.


\subsection{Stacking of the patchy screening effect}

In analogy with kSZ estimators, small-scale, real-space estimators have been proposed, with the potential of allowing patchy screening analyses to be performed with the same pipelines as kSZ. Here we use a similar estimator to v1 of \cite{2024arXiv240113033C, 2024PhRvD.109j3539S}:
\begin{equation}
\label{eq:tau_est}
\hat \tau_{\rm patchy} (\theta_d) = \frac{-\sum_i \mathrm{Sign}[T_i^\mathrm{lo}] T_i^\mathrm{hi}(\theta_d)}{N_g\langle |T^\mathrm{lo}|\rangle },
\end{equation}
where the sum and averaging is done over all $N_g$ galaxies in the sample. We note that unlike Ref.~\cite{2024arXiv240113033C} where $T^{\rm lo}_i$ is allowed to vary as a function of radius for each object, we take the sign of the low-pass filtered map at the pixel corresponding to the galaxy right ascension and declination. We find that this choice makes a negligible difference and prevents accruing a bias in some fringe cases. 
 
Similarly to the kSZ effect, this form of the estimator (that depends on the large-scale sign) is in principle immune to foreground emission such as tSZ and CIB. To further suppress the amount of residual foreground contamination to the final profiles, we randomly downsample the galaxy samples such that the number of positively and negatively weighted regions is equal. We also apply a cut on the amplitude of the low-pass-filtered temperature of $|T^{\rm lo}| > 40 \, \mu{\rm K}$. This is done since when that amplitude is small, the value of $T^{\rm lo}$ might be dominated by foregrounds (that can change the sign of $T^{\rm lo}$) rather than the primary CMB, which is what is needed for this measurement.

\subsection{Covariance matrix}

We estimate the covariance matrix for the stacked profiles by calculating the covariance between the radial bins per galaxy of $\tau_{\rm kSZ}$ and $\tau_{\rm patchy}$ and then simply dividing by $\sqrt{N_g}$. We note that because this calculation assumes that the galaxy profiles are independent of each other (and we know that many of the cutouts are overlapping), this covariance ends up underestimating the true covariance by around 10\% as demonstrated in Ref.~\cite{2024arXiv240113033C} when comparing with simulations. Since the main goal of this study is to compare the two measurements (kSZ and patchy screening) for the same sample of objects, we are less preoccupied with this effect as long as it affects both measurements equally.

\subsection{Simulations}

To put our findings into context, we compare the profiles we measure via ACT and DESI with the profiles measured in the hydrodynamical simulation IllustrisTNG. To this end, we construct 2D maps from the simulations of the kSZ or the optical depth, then convolve them with a 1.4 arcmin beam, and finally apply filtering as described in Section~\ref{sec:filter}.


The IllustrisTNG galaxy formation model features primordial and metal line cooling, a subgrid model for star formation and the interstellar medium, mass return from stars via active galactic nuclei (AGN) and supernovae, an effective model for galactic winds, as well as a model for the formation, growth, and feedback from supermassive black holes \citep{Weinberger2017,Pillepich2018}. At a baryonic mass resolution of $1.1\times 10^7 \mathrm{M_\odot}$, the highest-resolution IllustrisTNG simulation TNG300-1 reproduces well many properties of observed galaxies and galaxy clusters.

IllustrisTNG outputs a number of useful quantities: gas cell mass ($m$), electron abundance ($x$), and internal energy ($\epsilon$). Assuming a primordial hydrogen mass fraction of $X_H=0.76$, we compute the volume-weighted electron number density, $n_{\rm e}$, for each gas particle $i$ as
\begin{equation}
V_i n_{{\rm e},i} = x_i m_i \frac{X_H}{m_p}
\end{equation}
where $m_p$ the proton mass and $k$ is the Boltzmann constant. We then compute the 2D maps of the kSZ and the optical depth by binning the gas particles into a (10000, 10000) grid, so that the optical depth in cell $j$ is given by:
\begin{equation}
\tau_j = \sigma_T A_j^{-1} \sum_{i \in A_j} V_i n_{{\rm e},i} ,
\end{equation}
and the momentum of the electron density is:
\begin{equation}
b_j = \sigma_T A_j^{-1} \sum_{i \in A_j} V_i n_{{\rm e},i} v_i/c,
\end{equation}
where $A_j$ is the area of each grid cell (of size 0.03 Mpc), $\sigma_T$ is the Thomson cross section, and $c$ is the speed of light.

\section{Results}
\label{sec:results}

In this section, we present our main findings of comparing the kSZ and patchy screening signals measured on both data and simulations.

\subsection{Patchy screening and kSZ comparison}

\begin{figure}[h]
    \centering
    \includegraphics[width=0.45\textwidth]{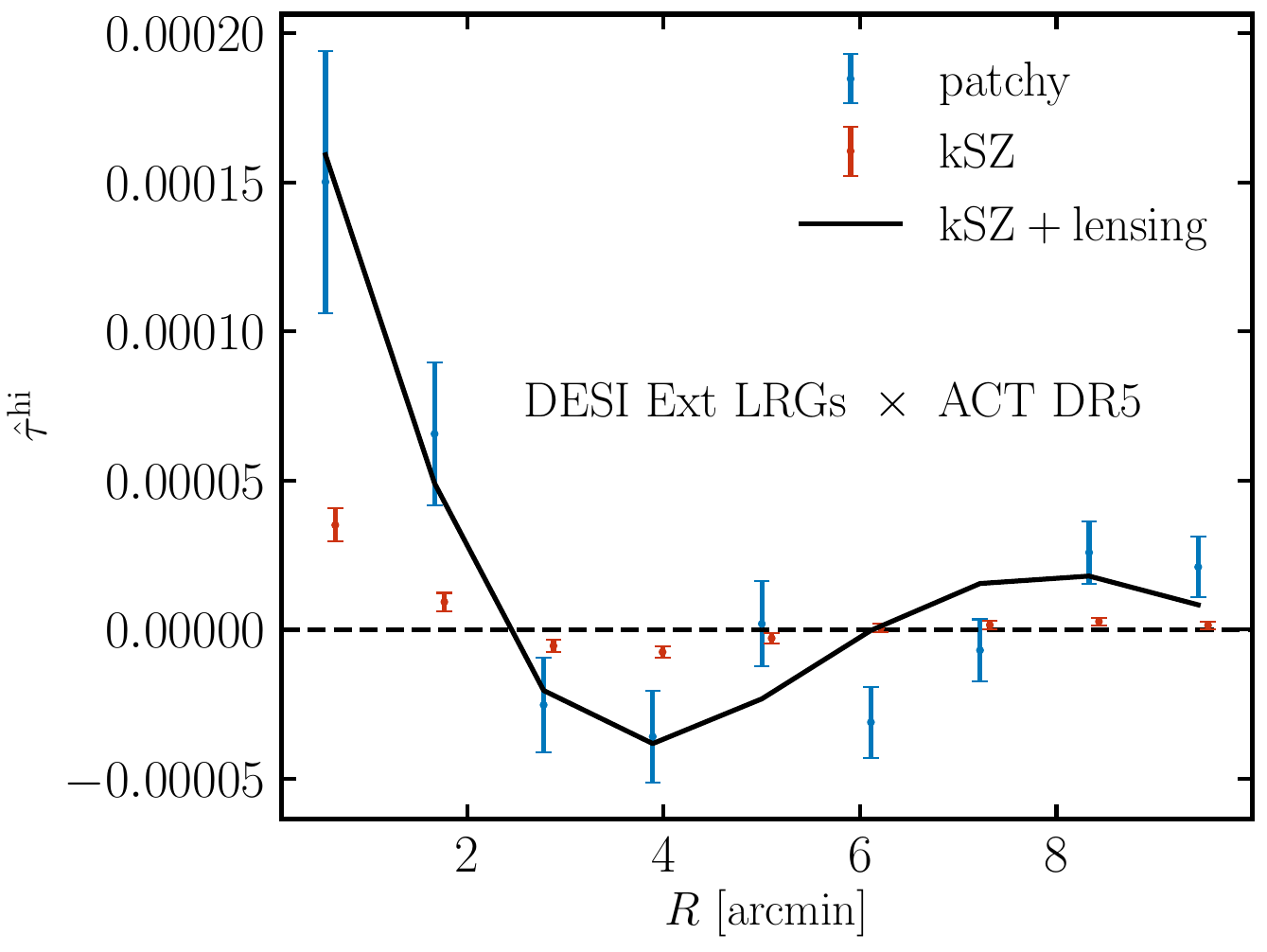}
    \caption{Estimates of the optical depth of DESI LRG groups using the two estimators defined in Eq.~\ref{eq:kSZ_est} and Eq.~\ref{eq:tau_est} and applying them to the ACT DR5 combined f090 and f150 maps. 
    We show that the two estimates of $\tau_{\rm hi}$ are in very good agreement if we take into consideration the contribution from CMB lensing to the `patchy screening' estimator (i.e., Eq.~\ref{eq:tau_est}), which makes up about two thirds of the total signal. The lensing contribution is assessed using the $N$-body simulation, \textsc{AbacusSummit} (see Section~\ref{sec:results}, for more details). A noteworthy feature is that the high-pass filtering of the CMB map makes the profiles with high, low and no baryonic feedback look very similar to each other (see also Ref.~\cite{2024arXiv240113033C}), rendering the comparison of profile shapes at high significance somewhat challenging given the signal-to-noise of the data.}
    \label{fig:tau_patchy_ksz}
\end{figure}

In Fig.~\ref{fig:tau_patchy_ksz}, we compare the signal measured from both effects, `patchy screening' and kSZ, on DESI LRG groups by applying the two estimators defined in Eq.~\ref{eq:kSZ_est} and Eq.~\ref{eq:tau_est} to the ACT DR5 combined f090 and f150 maps. 
We do so by filtering the f090 single-frequency map by the ratio of the two beams such that it becomes effectively rebeamed to the same beam size as the f150 map.
The significance of detection using the two estimators is 7.2$\sigma$ and 4.1$\sigma$, respectively, and the measured values of $\chi^2$ are 52.5 and 29.9, with 9 degrees-of-freedom. We obtain the signal-to-noise ratio as follows:
\begin{equation}
    {\rm SNR} = \sqrt{\chi^2_{\rm null} - \chi^2_{\rm best-fit}},
    \label{eq:SNR_null}
\end{equation}
where the $\chi^2$ statistic is defined as
\begin{equation}
    \chi^2_{\rm model} = (d - m)^\top {\rm C}^{-1} \hspace{0.1 cm} (d - m).
    \label{eq:chi_squared}
\end{equation}
In the case of $\chi^2_{\rm null}$, $m = 0$, whereas for the best-fit model, we adopt a rescaling of the simulation curves detailed in App.~\ref{app:sim_test} with a single free parameter for the amplitude, i.e. $m = A \ s$, where $s$ is the simulation curve of galaxies with a mean mass of $10^{13.3} \ M_\odot/h$ \citep{Yuan:2023ezi}. We note that because of the high-pass filtering, most of the information regarding the amount of feedback, halo mass, satellite fraction, and beam is contained in the amplitude rather than the shape of the curve. This is different from the CAP filter, for which the shape and amplitude convey independent information, as will be discussed in the next section.

To convert the kSZ signal measured on the temperature maps into optical depth, we do as follows:
\begin{equation}
    \tau_{\rm kSZ}^{\rm hi} = T_{\rm kSZ}^{\rm hi}/T_{\rm CMB} \  c/v_{\rm rms} \frac{1}{r} ,
\end{equation}
where the rms of the sample is assumed to be 300 km/s, and the cross-correlation coefficient between the true and reconstructed velocity is $r = 0.25$, which we verify in simulations for a photometric LRG-like sample in DR9 with no photo-$z$ cleaning performed on it as done in Ref.~\cite{2024arXiv240707152H}. 

We see that the amplitude of the kSZ signal appears to make up one third of what the estimator in Eq.~\ref{eq:tau_est} yields, and we attribute the difference between the two to weak lensing of the CMB giving a large contribution to the patchy screening estimator as we discuss next. This significantly lowers the detection significance of 4.1$\sigma$ and therefore, we quote an upper bound instead.

To put an upper bound on the optical depth detected through the `patchy screening' effect, we take the best-fit value of the amplitude parameter $A$ (see Eq.~\ref{eq:chi_squared}), which minimizes the difference between data and model. We note that as we are interested in the pure $\tau$ signal, here we work with the lensing-free curve, which can either be obtained by subtracting the lensing contribution from the `patchy screening' curve or by directly using the kSZ profile. The final result is virtually unchanged. The upper limit is then given by
\begin{equation}
A + N \sigma_A = A+\frac{N}{\sqrt{d^\top \mathrm{C}^{-1} d}}    ,
\end{equation}
where $\sigma_A$ is the variance of $A$ and $N = 1.64$ for a 95\% confidence level. We find that $A = 0.54$ and $\sigma_A = 0.61$. To convert this into a physically meaningful quantity, we note that the mean optical depth for the sample used in the theory prediction (mean halo mass of $10^{13.3} \ M_\odot/h$) is $\approx$ 1.6 $10^{-4}$, including the one- and two-halo terms \citep{2023MNRAS.526..369H}. Thus, our measurement places a 95\% upper bound on the optical depth of the Extended DESI LRG sample of $\tau <$ 2.5 $10^{-4}$.

\subsection{CMB lensing contribution to the stacked patchy screening estimator}
CMB lensing couples large and small scales, generally biasing patchy screening estimators. This was first pointed out by \cite{2011arXiv1106.4313S}, who noted that the lensing bias is generally larger than the signal itself, and proposed ``lensing hardening'' as a potential solution, which is noted to lead to a non-negligible noise cost. The lensing contribution to real-space estimators has not yet been quantified (to our knowledge), and as discussed below, it is expected to be very significant. We discuss the origin of the lensing bias and several mitigation strategies in a companion paper \citep{Sailer2024}.

The lensing contribution is estimated by adopting the publicly available convergence maps of \textsc{AbacusSummit} (namely, the all-sky `huge' simulation \texttt{AbacusSummit\_huge\_c000\_ph201}) and lensing a random Gaussian realization of the angular power spectrum, $C_\ell$, from \texttt{CAMB} using \texttt{pixell} \cite{2023MNRAS.525.4367H}. We then stack at the location of the \textsc{AbacusSummit} halos on the light cone \cite{2022MNRAS.509.2194H} in the redshift range corresponding to the DESI LRGs: $z = 0.45$ to 1.  Note that the generated lensing maps do not contain a $\tau$ signal and can thus be independently added to a pure $\tau$ curve, as done in Fig.~\ref{fig:tau_patchy_ksz}. Finally, we select a low-mass threshold, so that the mean halo mass roughly matches that of the LRG host halos, $\log M \approx 13.3$ in units of $M_\odot/h$. We perform an additional test to check that we recover the correct relative amplitude between the lensing and the $\tau$ signals and reaffirm our finding (App.~\ref{app:sim_test}). 

Naively, one might expect that the matter profiles measured by lensing would be different from the gas profiles that make up a fraction of the total `patchy screening' signal, especially on the outskirts of groups and clusters, where we expect the gas to be pushed out compared with the dark matter. However, we note that because of the high-pass filtering of the CMB map, information on these larger scales is largely erased, making the shapes of the profiles with high, low and no baryonic feedback look virtually indistinguishable from each other (see also Ref.~\cite{2024arXiv240113033C}) given the signal-to-noise of the data. 

Instead, the amplitude now contains information about the gas distribution (provided lensing is accounted for in the `patchy screening' case) and thus feedback, while also being sensitive to the satellite fraction and the mean halo mass of the sample. This sensitivity comes from the fact that stacking at the location of satellite galaxies leads to more extended profiles compared with the case of stacking at the center of the halos (see Fig.~\ref{fig:CAP_hi}), which then curbs the power at high-$\ell$. We note that the beam of a CMB experiment has the same effect, as it puffs out the mass distribution. 

Similarly, at fixed halo mass (and we remind the reader that both the $\tau$ and lensing signals are proportional to halo mass), galaxy groups with little feedback would have a higher amplitude than galaxy groups with a large amount of feedback. As we increase the halo mass (keeping the baryonic properties and satellite fraction fixed), the amplitude increases proportionally to the halo mass until the size becomes comparable to the size of the high-pass filter threshold applied to the CMB map, at which point some of the amplitude will be lost, as it leaks to larger scales than allowed by the filtering.

However, if one can pinpoint the halo occupation distribution of the sample, then the gas feedback can be constrained. Additionally, provided that the effect of CMB lensing is well understood, the comparison between the kSZ $\tau^{\rm hi}$ amplitude and that of the bias-hardened `patchy screening' $\tau^{\rm hi}$ could yield a powerful probe of the typical velocities of the galaxy sample host halos and can also be used to calibrate $r$ independently of $N$-body simulations \cite{2024PhRvD.109j3533R,2024PhRvD.109j3534H}, providing a long-sought solution to the `optical-depth degeneracy' problem of disentangling astrophysical quantities ($\tau$) from cosmological ones ($v_{\rm rms}$) \citep{2021PhRvD.103f3518A}.

\subsection{Size of the baryonic feedback}

\begin{figure}[h]
    \centering
    \includegraphics[width=0.45\textwidth]{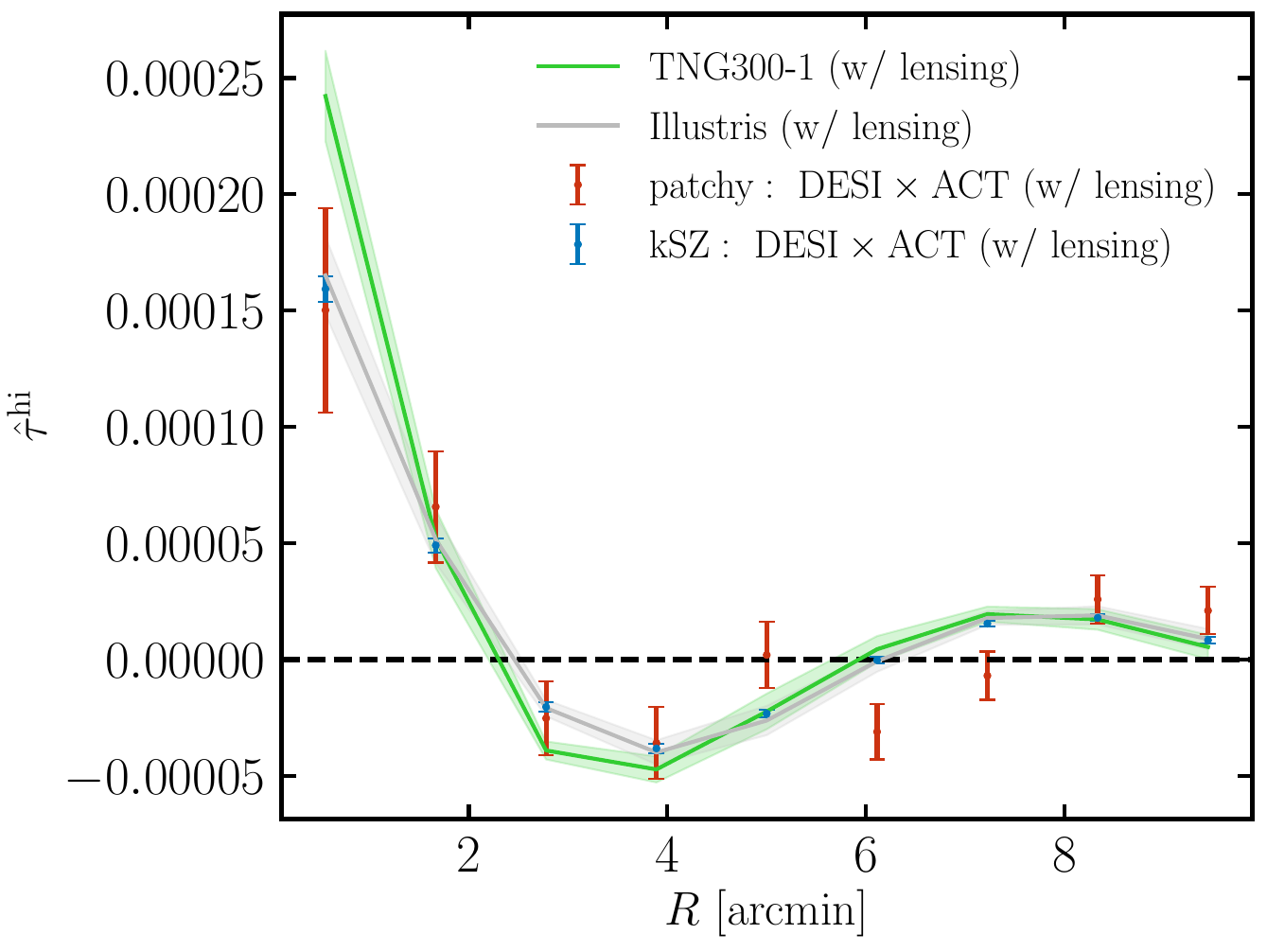}
    \caption{Optical depth signal in the `patchy screening' measurement from DESI LRG $\times$ ACT DR5, obtained by stacking via the estimator in Eq.~\ref{eq:tau_est}. To make the comparison with simulations and the kSZ, we add the lensing contribution to the kSZ and simulations at the mean halo mass of the galaxy sample, $\log M \approx 13.3$, using the \textsc{AbacusSummit} weak lensing maps. The measured $\tau$ profile has a lower amplitude compared with TNG300-1, implying that the true baryonic feedback is stronger than the one predicted in TNG300-1. 
    The profiles are more akin to the Illustris $\tau$ profile, implying that the amount of feedback Illustris predicts, which is large compared with the majority of hydro simulations, reflects more accurately the extent to which gas is expelled in the Universe through processes such as AGN feedback.}
    \label{fig:tng_illustris_tau}
\end{figure}

First, we inspect the optical depth signal in the `patchy screening' measurement from DESI LRG $\times$ ACT DR5, obtained by stacking via the estimator in Eq.~\ref{eq:tau_est}. This is shown in Fig.~\ref{fig:tng_illustris_tau}. For all other samples (i.e., the kSZ data measurement and the $\tau$ simulation curves), we add the lensing contribution at the mean halo mass of the galaxy sample, $\log M \approx 13.3$, using the \textsc{AbacusSummit} weak lensing maps. The two hydro simulation curves are obtained by painting gas properties onto the \textsc{AbacusSummit} light cone maps via the method outlined in App.~\ref{app:sim_test}. We see that the $\tau$ profile measured in the data has a much lower amplitude compared with TNG300-1, implying that the true baryonic feedback is stronger than the one predicted in TNG300-1. 
In fact, we see that the data appears to be much more similar in shape and amplitude to the Illustris $\tau$ profile, implying that the amount of feedback Illustris predicts, which is large compared with the majority of hydro simulations, reflects more accurately the extent to which gas is expelled in the Universe through processes such as AGN feedback. Reassuringly, we find great consistency with previous kSZ measurement analyses \citep{2021PhRvD.103f3514A,2024arXiv240707152H,2024MNRAS.534..655B,2024arXiv241019905M,2024arXiv241203631H}, providing further evidence for strong baryonic feedback in the LRG sample. 

We note that while the transfer function method (App.~\ref{app:sim_test}) does an extremely good job of mimicking the gas distribution in a hydro simulation at scales larger than $R \sim 1 \ {\rm Mpc}/h$ ($k \lesssim 1 \ h/{\rm Mpc}$ in Fourier space) \citep{Liu2024}, on smaller scales, where the cross-correlation coefficient between the dark matter density and $\tau$ becomes significantly smaller than one, the painted-on gas density will deviate from the gas density in the hydro simulation. We note that this affects the first couple of radial bins of the `painted' Illustris curve, which end up having higher amplitude compared with the `hydro' Illustris curve (we have verified that to be the case at lower masses, for which we have a representative sample of halos in the hydro simulation). 

\subsection{Significance of the CAP filter}

\begin{figure}[h]
    \centering
    \includegraphics[width=0.45\textwidth]{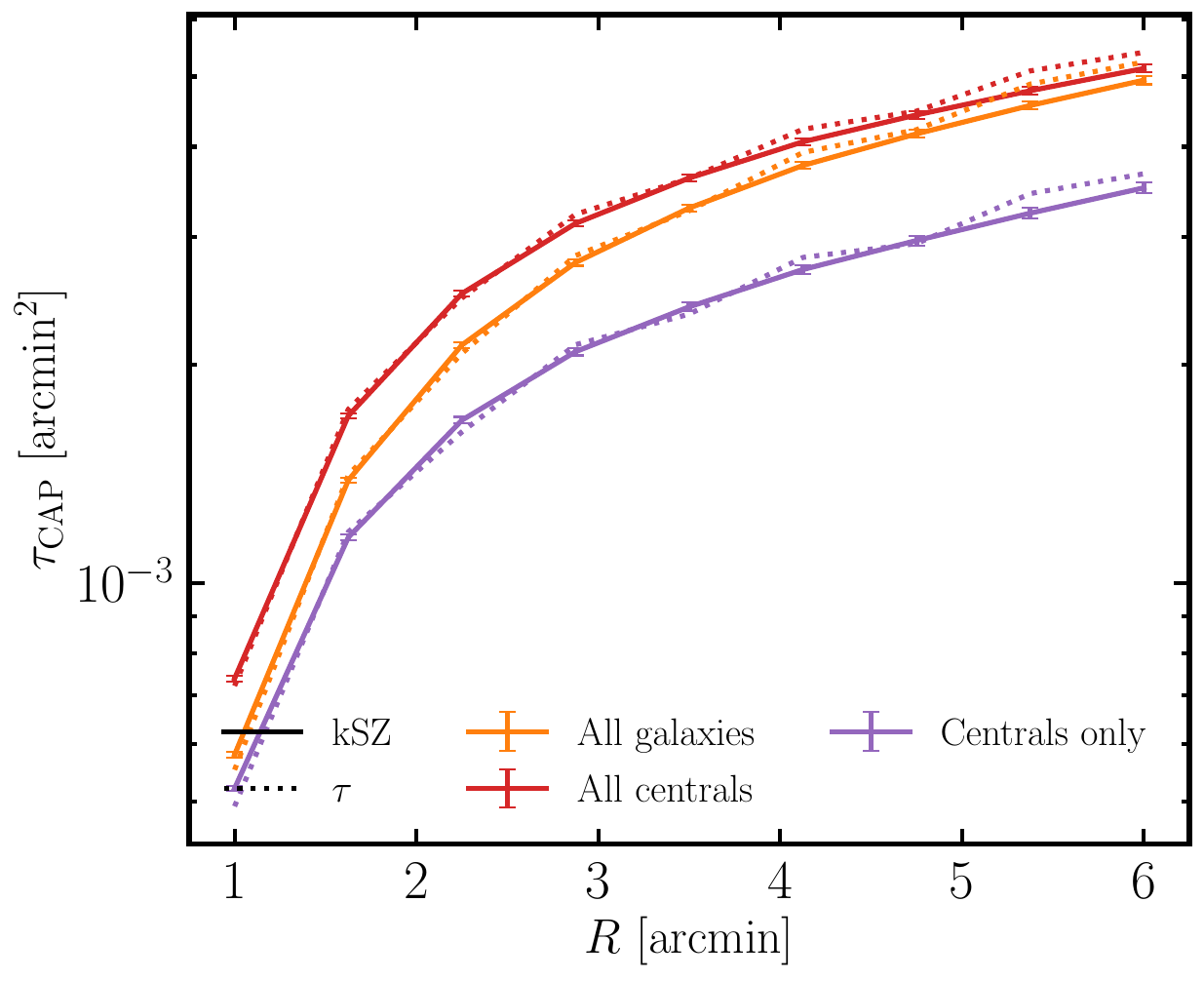}
    \includegraphics[width=0.45\textwidth]{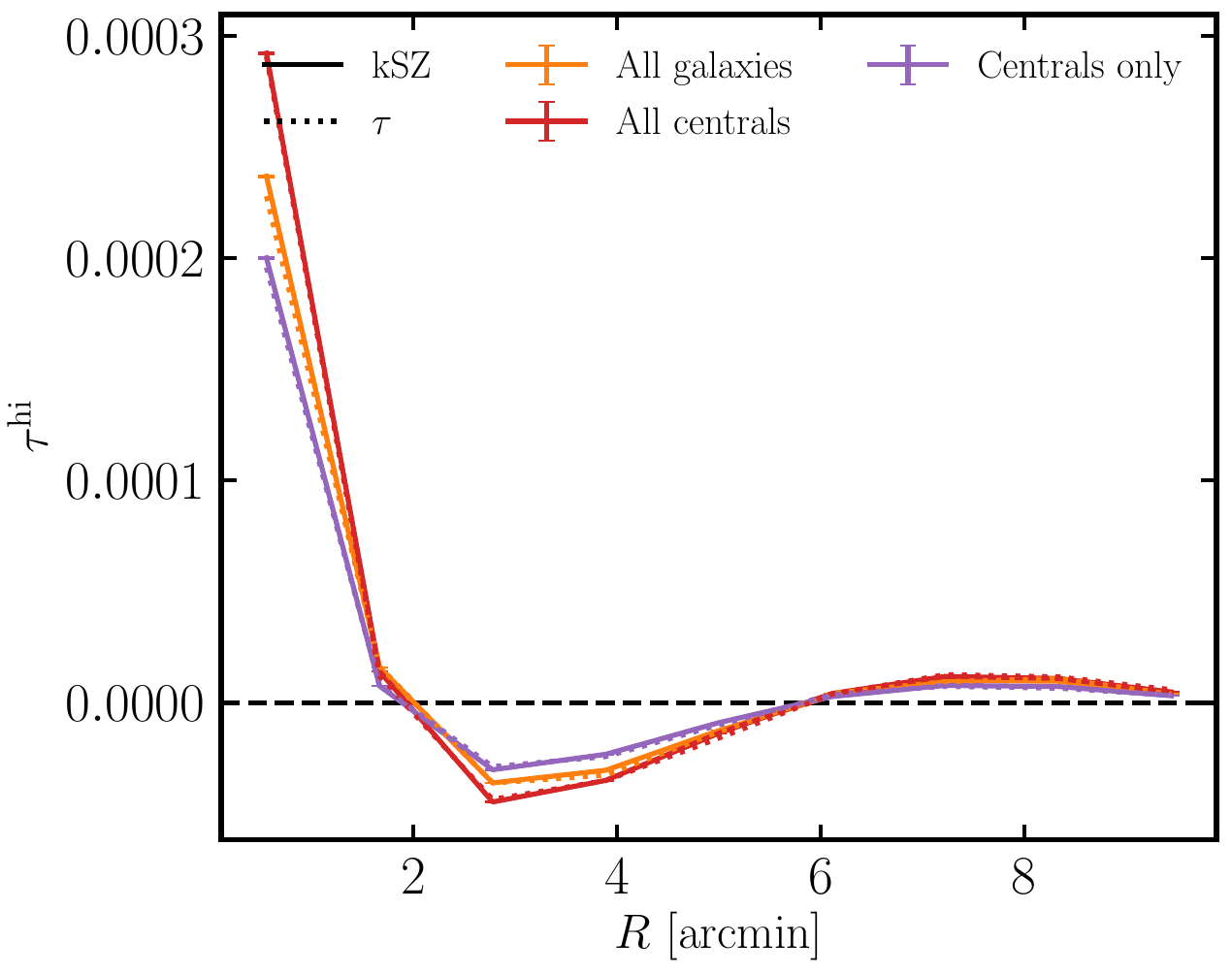}
    \caption{CAP (\textit{top}) and high-pass-filtered (\textit{bottom}) profiles measured in IllustrisTNG (TNG300-1) around high-mass galaxies from 2D maps of the optical depth, $\tau$ (dotted), and the kSZ effect (solid)
    at $z = 0.5$. We note that the profiles are not convolved with the ACT beam and that the maps contain no primary CMB or lensing contributions. The units of the signal in the top panel are steradians, as the CAP filter measures the difference between the integrated signal in a disk and in a ring around it (see Eq.~\ref{eq:ap}). The quantity on the $y$-axis can thus be thought of as a cumulative baryon profile, whereas $\tau^{\rm hi}$ in the bottom panel is unitless, as it corresponds to the mean high-pass filtered optical depth. The three colors correspond to stacking on the default LRG sample (`All galaxies'), stacking on only the centrals belonging to that sample, or around 80\% of the galaxies (`Centrals only'), and stacking on all galaxies, but at the location of their host halo centers to avoid miscentering (`All centrals'). Most noteworthy is the fact that the kSZ and $\tau$ lie on top of each other, suggesting that they measure the same underlying quantity of optical depth. 
    }
    \label{fig:CAP_hi}
\end{figure}

In the top panel of Fig.~\ref{fig:CAP_hi}, we show the CAP profiles of high-mass galaxies (i.e., with $log M_\ast > 10.45$) 
measured in IllustrisTNG (TNG300-1) obtained by stacking on 2D maps of the optical depth and the kSZ effect at $z = 0.5$. We show the results for the default LRG sample (`All galaxies'), the centrals belonging to that sample (`Centrals only'), and the halo centers for all galaxies (`All centrals'). This quantity can be thought of as a cumulative baryon profile. We note that these maps do not have the primary or lensed CMB signal. Furthermore, on very small scales ($\sim$1 arcmin), the transfer function method breaks down, as the cross-correlation coefficient between the dark matter field and the gas field in the Illustris simulation becomes small ($r \ll 1$).

Unsurprisingly, at large aperture, the `All centrals' and `All galaxies' curves reach the same value of total gas mass. In further comparing the `All galaxies' sample with the `All centrals,' we see that the main difference is that `All centrals' appears slightly flatter. This is due to the fact that the non-zero satellite fraction of the `All galaxies' sample leads to miscentering, which artificially makes the stacked profiles more extended. The amplitude, as expected, is the same as the mean halo mass remains unchanged. 

On the other hand, the `Centrals only' sample has slightly lower amplitude as when stacking on centrals only, we have discarded satellite galaxies, which are more likely to reside within massive groups and clusters, and have thus lowered the mean halo mass of the sample. That is also the main reason for the difference in shape: namely, the mean virial radius of that sample is smaller, making the profiles flatter at fixed apparent size on the sky. The error bars here are calculated as the standard deviation of the measured curves from the sample divided by the square root of objects in the sample. The CAP filter cannot be applied to the `patchy screening' estimator due to leakage of the large-scale CMB fluctuations (see also App.~\ref{app:snr}), so when studying that effect, we need to adopt high-pass filtered profiles.

In the bottom panel, analogously to the CAP case, we study the high-pass filtered profiles. Since in this case, we are taking the average of the signal in rings of increasing distance, the highest point of the profile sits in the center of the halo, as expected, where most of the gas resides. The dip below zero and then rise of the signal at large radii is due to the ringing effect at imposing the cutoff on the 2D map. As discussed previously, the amplitude of the high-pass filtered signal is lowered in the presence of satellite galaxies, as the density profile becomes more puffed out and the high-pass filter takes away some of that power on large scales. 

Reassuringly, for both filters (CAP and high-pass) the kSZ and $\tau$ lie on top of each other, indicating that they measure the same underlying quantity of optical depth. The role of the CAP filter is important, as it removes the uncorrelated part of the signal, which dominates the optical depth map, leaving the contribution of the targeted galaxies only. Similarly, the high-pass filter gets rid of the largest-scale contributions both from the primary CMB and also from random structure along the line-of-sight. In the case of the kSZ, that is less of an issue as due to the velocity decorrelation along the line-of-sight, the signal picks up predominantly the contribution from the targeted object \cite{2023MNRAS.526..369H}. 


\section{Discussion and conclusions}
\label{sec:conclusions}

In this paper, we pursue a comparison between the stacked measurements of the gas density using the kinematic Sunyaev-Zel'dovich effect and the `patchy screening' effect. Both of these probes are sensitive to the gas distribution and can thus help us inventory the missing baryons in the Universe. In principle, the patchy screening effect is about 25 times smaller than the kSZ effect, but it can be applied to 2D surveys without any knowledge of the line-of-sight velocity field, which is necessary for the kSZ effect. As such, the combination of these two effects provides a powerful way of disentangling the astrophysical quantity, optical depth, from the cosmological one, peculiar velocity, breaking the so-called `optical depth degeneracy' and yielding an independent measurement of the line-of-sight velocity.

Here, we work with the Extended LRG sample of the DESI imaging survey DR9, which has previously been used for measuring the gas density profiles using the kSZ effect. We then high-pass filter the CMB temperature map and apply the estimator proposed in v1 of Ref.~\cite{2024arXiv240113033C} (Eq.~\ref{eq:tau_est}) and the kSZ estimator (Eq.~\ref{eq:kSZ_est}) to consistently compare the two. Our main findings are summarized as follows:
\begin{itemize}
    \item In Fig.~\ref{fig:tau_patchy_ksz}, we present a high-significance detection of both effects: namely, we measure the kSZ signal at 7.2$\sigma$ and the `patchy screening' signal at 4.1$\sigma$, the latter of which turns out to be contaminated by CMB lensing. 
    \item We show that the `patchy screening' estimator (Eq.~\ref{eq:tau_est}) receives a large contribution from CMB lensing, which makes up about two thirds of the measured signal for the chosen sample. While this paper uses simulations to estimate the effect of lensing, in an accompanying paper \citep{Sailer2024}, the focus is on deriving an analytical expression to account for this contamination. 
    \item As explored in App.~\ref{app:snr}, after removing the lensing portion, the signal-to-noise for $\tau$ coming from `patchy screening' of 4.1$\sigma$ is lowered significantly, 
so we can quote an upper bound of $\tau < 2.5 \ 10^{-4}$ for our sample, which we assume to have a mean optical depth of $\tau \approx 1.6 \ 10^{-4}$. We note that the effect is still measurable with future surveys such as the Vera Rubin Observatory (LSST) \cite{2019ApJ...873..111I}.
    \item By performing stacked measurements on hydro simulations of pure kSZ and pure $\tau$ maps (i.e., without the inclusion of other CMB primary and secondary effects), we are reassured that the kSZ estimator (Eq.~\ref{eq:kSZ_est}) is indeed measuring the correct underlying quantity of optical depth, $\tau$. 
\end{itemize}

In the future, we aim to explore ways of combining these two potent measurements to acquire a detailed picture of the baryon distribution in the Universe. The ultimate constraints on the thermodynamics of galaxy groups and clusters and by extension the effects on the small-scale matter power spectrum will come from obtaining a multi-scale multi-wavelength view of the Universe through additional powerful probes such as tSZ, X-rays and Fast Radio Bursts (FRBs).

\acknowledgements

We thank Matt Johnson for a useful discussion during the preparation of this manuscript.
N.S. and S.F. are supported by Lawrence Berkeley National Laboratory and the Director, Office of Science, Office of High Energy Physics of the U.S. Department of Energy under Contract No.\ DE-AC02-05CH11231.

\appendix

\section{Simulation test of patchy screening and the effect of CMB lensing}
\label{app:sim_test}

\begin{figure}[h]
    \centering
    \includegraphics[width=0.45\textwidth]{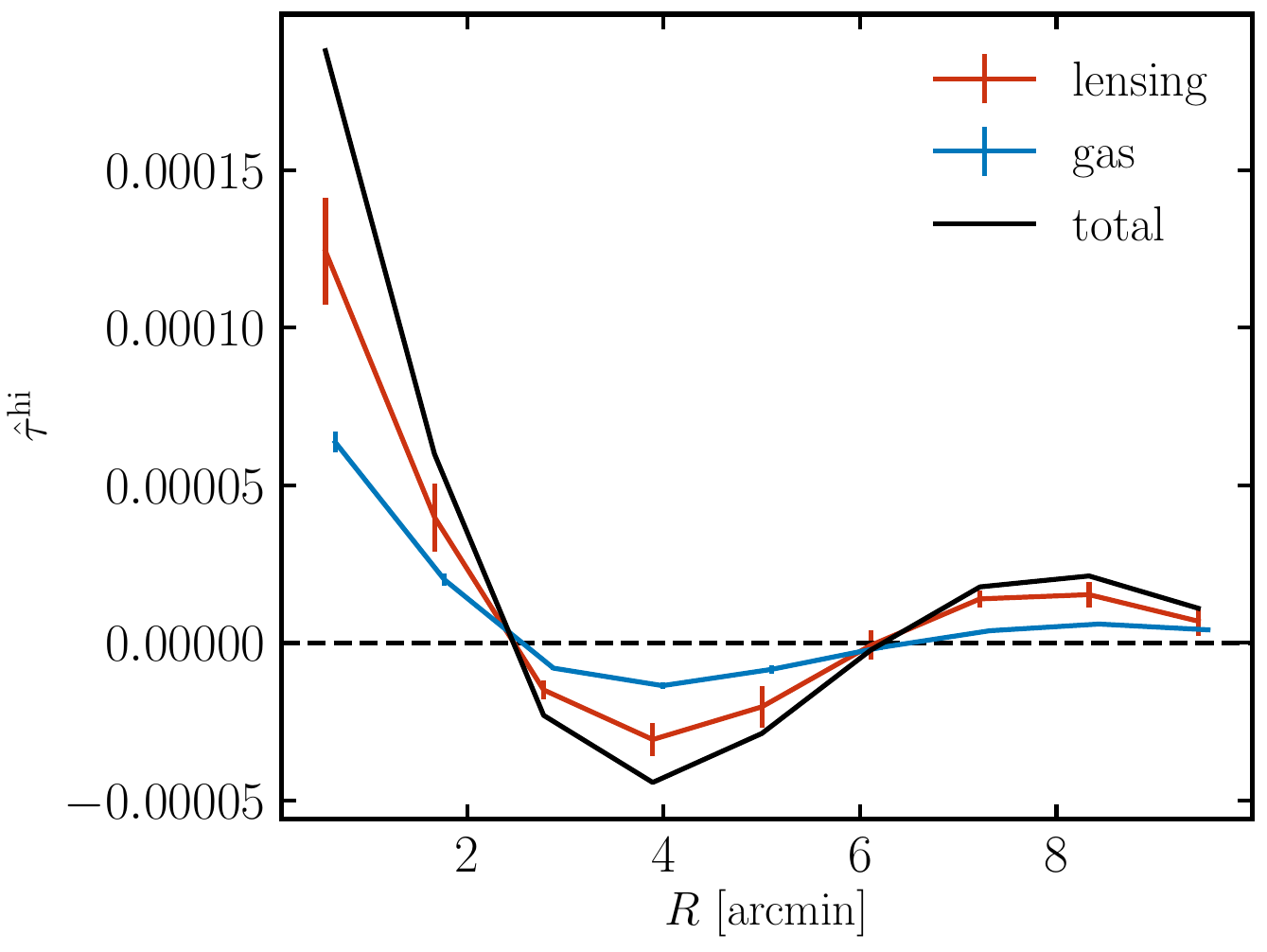}
    \caption{Simulated high-pass-filtered profiles of the `patchy screening' effect as implemented in Eq.~\ref{eq:tau_est}. Maps are generated by applying the `transfer function method' to the AbacusSummit huge box density maps between $z = 0.45$ and 1 to match as closely as possible the LRG sample redshift support (see the text in App.~\ref{app:sim_test} for more details). Stacks are performed on all halos in that redshift range above $10^{13.07}$ such that the mean halo mass matches that of the LRG sample, $10^{13.3}$ in units of $M_\odot/h$. 
}
    \label{fig:gas_lensing_sim}
\end{figure}

In addition to the lensed maps, we also create `patchy screening'-only maps by generating a random Gaussian relization from the lensed $C_\ell$ and then multiplying the thus-obtained CMB temperature map by $\exp[-\tau]$. To get the $\tau$ map on the full sky, we adopt the transfer function technique presented in \citep{2018ApJ...853..121P,Liu2024}, for which the authors show that multiplying the matter field of an $N$-body simulation by the transfer function:
\begin{equation}
    T(k) = \sqrt{\frac{P_{\tau\tau}}{P_{\rm mm}}}
\end{equation}
in Fourier space leads to a new field $\hat \tau$ that matches the true $\tau$ field of a hydro simulation at the field level down to very small scales. As noted previously \citep{2024arXiv240707152H}, Illustris at $z = 0.5$ provides a good fit to the optical depth profiles observed for the LRGs. Thus, we adopt the transfer function from the Illustris simulation when constructing the theory model used in Eq.~\ref{eq:chi_squared}. 

Next since we work with the HEALPix density sheets rather than the 3D density fields \citep{2005ApJ...622..759G}, we convolve each sheet with a transfer function in harmonic space evaluated as $T(\ell = k \chi)$, where $\chi$ is the comoving distance to the sheet. This finding holds across different feedback models and can be quantified via the cross-correlation coefficient, $r(k)$, between $\tau$ and $\hat \tau$, which remains above 0.95 until $k \approx 1.25 \ h/{\rm Mpc}$ for Illustris \cite{2015A&C....13...12N}, used in this study. 

The result of the comparison is shown in Fig.~\ref{fig:gas_lensing_sim}. We see that the contribution from the gas portion of the signal, i.e. $\tau$, makes up about a third of the signal, with the lensing contributing about two thirds. The amplitude of the signal is likely somewhat exaggerated due to the fact we are not accounting for miscentering in this test performed on halos, and on very small scales ($\sim$1 arcmin), the transfer function method breaks down, as the cross-correlation coefficient between the dark matter field and the gas field in the Illustris simulation becomes small, i.e. $r(k) = P_{\tau m}/\sqrt{P_{\tau \tau} P_{m m}} \ll 1$.

\section{Comparison of the signal-to-noise}
\label{app:snr}

\begin{figure}[h]
    \centering
    \includegraphics[width=0.45\textwidth]{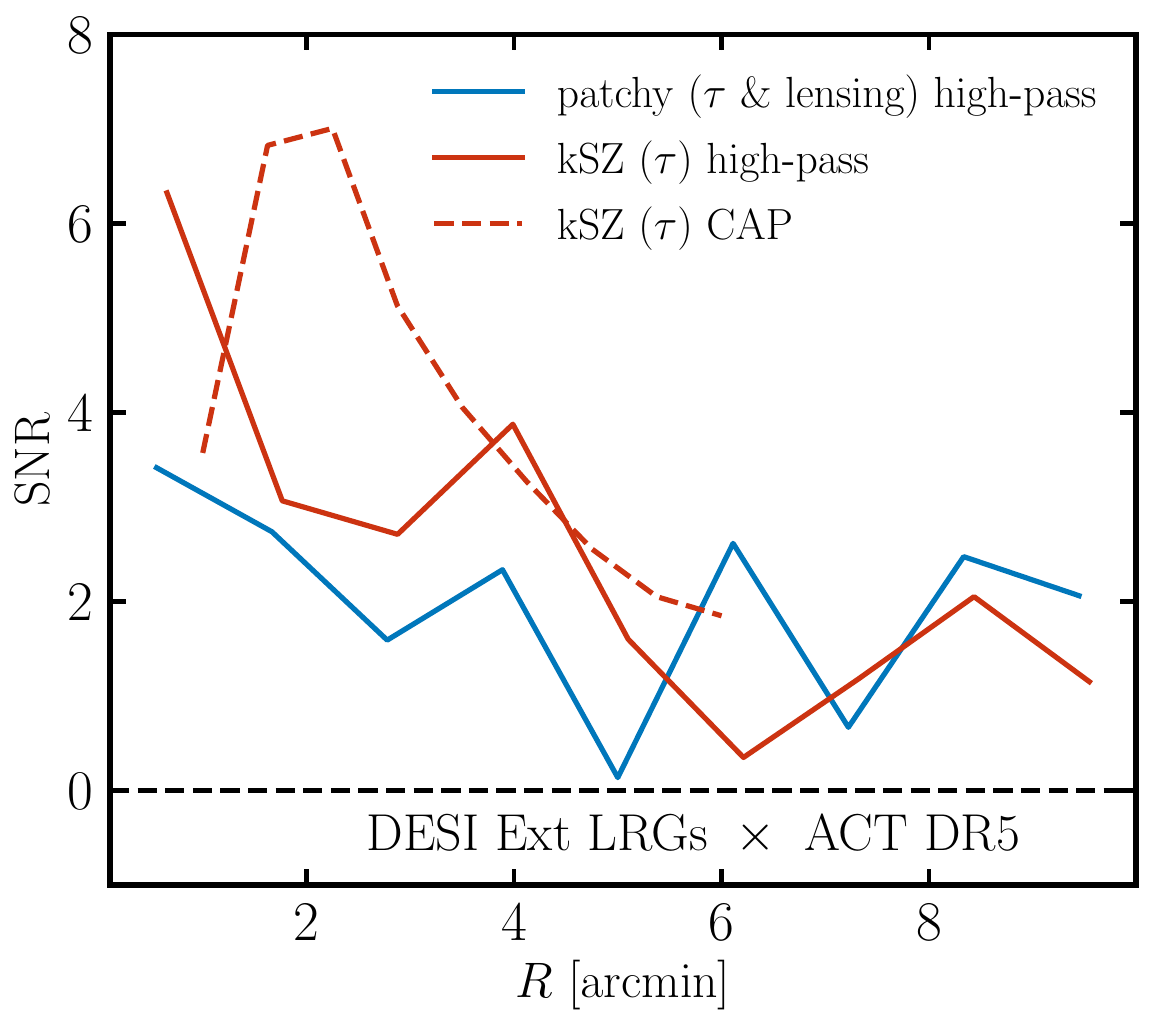}
    \caption{Comparison of the signal-to-noise per radial bin for our measurements using ACT and DESI data of the `patchy screening' signal, resulting from $\tau$ and lensing, and the kSZ effect, resulting from $\tau$ after reconstructing the line-of-sight velocity field.
}
    \label{fig:tau_snr}
\end{figure}

In Fig.~\ref{fig:tau_snr}, we compare the signal-to-noise per radial bin for our measurements using ACT and DESI data of the `patchy screening' signal, resulting from $\tau$ and CMB lensing, and the kSZ effect, resulting from $\tau$ after reconstructing the line-of-sight velocity field. We see that the majority of the signal-to-noise for the high-pass filtered quantities comes at small aperture and decreases towards large ones, where part of the signal has been filtered out. Similarly, the CAP-filtered quantity does not have a lot of sensitivity at large apertures, where the noise is dominated by the primary CMB. Most of its sensitivity comes at small apertures (2-4 arcmin), with the first radial bin being dominated by the ACT beam noise. We stress that the CAP filter cannot be applied to the `patchy screening' estimator due to leakage of the large-scale CMB fluctuations.

\bibliography{apssamp}

\end{document}